# NEW CONSTRAINTS ON COSMIC POLARIZATION ROTATION FROM B-MODE POLARIZATION IN COSMIC MICROWAVE BACKGROUND

Sperello di Serego Alighieri[1], Wei-Tou Ni[2] and Wei-Ping Pan[2]

[1]INAF-Osservatorio Astrofisico di Arcetri,
Largo Enrico Fermi 5, 50125 Firenze, Italy
*E-mail:* sperello@arcetri.astro.it

[2]Department of Physics, National Tsing Hua University,
Hsinchu, Taiwan 30013, Republic of China
*E-mail:* weitou@gmail.com , d9722518@oz.nthu.edu.tw

## ABSTRACT

STPpol, POLARBEAR and BICEP2 have recently measured the cosmic microwave background (CMB) B-mode polarization in various sky regions of several tens of square degrees and obtained BB power spectra in the multipole range 20-3000, detecting the components due to gravitational lensing and to inflationary gravitational waves. We analyze jointly the results of these three experiments and propose modifications of their analysis of the spectra to include in the model, in addition to the gravitational lensing and the inflationary gravitational waves components, also the effects induced by the cosmic polarization rotation (CPR), if it exists within current upper limits. Although in principle our analysis would lead also to new constraints on CPR, in practice these can only be given on its fluctuations $<\delta\alpha^2>$, since constraints on its mean angle are inhibited by the de-rotation which is applied by current CMB polarization experiments, in order to cope with the insufficient calibration of the polarization angle. The combined data fits from all three experiments (with 29% CPR-SPTpol correlation, depending on theoretical model) gives constraint $<\delta\alpha^2>^{1/2} <$ 27.3 mrad (1.56°) with r = 0.194 ± 0.033. These results show that the present data are consistent with no CPR detection and the constraint on CPR fluctuation is about 1.5°. This method of constraining the cosmic polarization rotation is new, is complementary to previous tests, which use the radio and optical/UV polarization of radio galaxies and the CMB E-mode polarization, and adds a new constraint for the sky areas observed by SPTpol, POLARBEAR and BICEP2.

## 1. INTRODUCTION

The BICEP2 collaboration (Ade et al. 2014b) has recently detected the cosmic microwave background (CMB) B-mode (tensor mode) polarization and finds an excess power around $\ell\sim80$ over the gravitational lensing expectation with a significance of more than 5σ, which they interpret as due to inflationary gravitational waves with a tensor-to-scalar ratio r = $0.20^{+0.07}_{-0.05}$. This result is in apparent contrast with the limit previously set by the Planck collaboration r<0.11 at 95% CL (Ade et al. 2014a). Three processes can produce B-mode polarization: (i) gravitational lensing from E-mode polarization (Zaldarriaga & Seljak 1997), (ii) local quadrupole anisotropies in the CMB within the last scattering region by large scale gravitational waves (Polnarev 1985) and (iii) cosmic polarization rotation (CPR)[1] due to pseudoscalar-photon interaction (Ni 1973; for a review, see Ni 2010). CPR is currently constrained to be less than about a couple of degrees by measurements of the linear polarization of radio galaxies and of the CMB (see di Serego Alighieri et al. 2011 for a review). However, if CPR exists within the current upper limits, it would produce a non-negligible B-mode CMB polarization. The BICEP2 Collaboration (Ade et al. 2014b) has not considered this latter component in their model. To look for new constraints on the CPR and into the robustness of the BICEP2 fit, we include the CPR effect in the model. TE, TB and EB correlations potentially give mean values of CPR angle <α>, while the contribution of CPR effects to B-mode power potentially gives <α>$^2$ plus the variations of the CPR angle squared <δα$^2$>. This method of constraining the CPR is new, is complementary to previous tests, and adds a new constraint for the sky area observed, although its application to current data is limited by the uniform angle de-rotation, which is applied to the measured CMB Q and U maps to compensate for insufficient calibrations of the polarization angle (Keating et al. 2013). We include in our analysis of the B-mode power spectra also the data available for $\ell$ >500, in particular from SPTpol (Hanson et al. 2013) and POLARBEAR (Ade et al. 2014c), in addition those from BICEP2 for smaller $\ell$.

In section 2, we review pseudoscalar-photon interaction, its modification on Maxwell equations and the associated electromagnetic propagation effect on polarization rotation. In section 3 we present the results of our fits. In section 4 we review and discuss various constraints on CPR and in section 5 we discuss a few issues and present an outlook towards the future.

---

[1] The CPR has also been inappropriately called Cosmological Birefringence, but we follow here the recommendation of Ni (2010).

## 2. PSEUDOSCALAR-PHOTON INTERACTION, MODIFIED PROPAGATION AND POLARIZATION ROTATION

Einstein Equivalence Principle (EEP) which assumes local special relativity is a major cornerstone of general relativity and metric theories of gravity. Special relativity sprang from Maxwell-Lorentz electrodynamics. Maxwell equations in terms of field strength $F_{kl}$ (***E***, ***B***) and excitation $H^{ij}$ (***D***, ***H***) do not need metric as primitive concept. Field strength $F_{kl}$ (***E***, ***B***) and excitation $H^{ij}$ (***D***, ***H***) can all be independently defined operationally (e. g. Hehl & Obukhov 2003). To complete this set of equations, one needs the constitutive relation between the excitation and the field in both macroscopic electrodynamics and in space-time theory of gravity:

$$H^{ij} = (1/2)\, \chi^{ijkl}\, F_{kl}. \qquad (1)$$

When Einstein Equivalence Principle is observed, this fundamental space-time tensor density is induced by the metric $g^{ij}$ of the form

$$\chi^{ijkl} = (1/2)\,(-g)^{1/2}\,(g^{ik}\,g^{jl} - g^{il}\,g^{kj}), \qquad (g \equiv (\det g^{ij})^{-(1/2)}) \qquad (2)$$

and the Maxwell equations can be derived from a Lagrangian density. In local inertial frame, the space-time tensor has the special relativity form $(1/2)\,(\eta^{ik}\,\eta^{jl} - \eta^{il}\,\eta^{kj})$ with $\eta^{il}$ the Minkowski metric. To study the empirical foundation of Einstein Equivalence Principle, it is crucial to explore how the experiments/observations constrain the general space-time tensor $\chi^{ijkl}$ to the GR/metric form.

Since both $H^{ij}$ and $F_{kl}$ are antisymmetric, $\chi^{ijkl}$ must be antisymmetric in $i$ and $j$, and $k$ and $l$. Hence the constitutive tensor density $\chi^{ijkl}$ has 36 independent components, and can be decomposed to principal part (P), axion part (Ax) and Hehl-Obukhov-Rubilar skewon part (Sk) (Hehl & Obukhov 2003). The skewon $^{(Sk)}\chi^{ijkl}$ part is antisymmetric in the exchange of index pair $ij$ and $kl$, satisfies a traceless condition, and has 15 independent components. The principal part and the axion (pseudoscalar) part constitute the parts that are symmetric in the exchange of index pair $ij$ and $kl$. Together they have 21 independent components. Axion (pseudoscalar field) part is totally antisymmetric in all 4 indices and can be expressed as $\varphi\, e^{ijkl}$ with $\varphi$ the pseudoscalar field and $e^{ijkl}$ the completely antisymmetric Levi-Civita symbol. The principal part then have 20 degrees of freedom.

In constraining the general space-time tensor from experiments and observations, we notice that EEP is already well-tested and can only be violated weakly. Hence, one can start with (2) adding a small general $^{(Sk)}\chi^{ijkl}$ to look for constraint on skewons first. From the dispersion relation it is shown that no dissipation/no amplification in propagation implies that the additional skewon field must be of Type II (Ni 2014). For Type I skewon field, the dissipation/amplification in propagation is proportional to the frequency and the CMB spectrum would deviate from Planck spectrum. From the high precision agreement of the CMB spectrum to 2.755 K Planck spectrum (Fixsen 2009), the Type I cosmic skewon field $|^{(SkI)}\chi^{ijkl}|$ is constrained to less than a few parts of $10^{-35}$ (Ni 2014). Generic Type II skewon field can be constructed from antisymmetric part of an asymmetric metric and is allowed.

EEP implies that photons with the same initial position and direction follow the same world line independent of energy (frequency) and polarization, i.e. no birefringence and no polarization rotation. This is observed to high precision for no birefringence and constrained for no polarization rotation. Since Type II skewon field in the weak field limit and axion (pseudoscalar) field do not contribute to the dispersion relation in the eikonal approximation (geometrical optics limit), the no-birefringence condition only limits the principal part space-time tensor $^{(P)}\chi^{ijkl}$ to

$$^{(P)}\chi^{ijkl} = \tfrac{1}{2} (-h)^{1/2}[h^{ik} h^{jl} - h^{il} h^{kj}]\psi, \qquad (3)$$

where the light cone metric $h^{ik}$ can be expressed in terms of $^{(P)}\chi^{ijkl}$ and Minkowski metric (Ni 1983a, 1984). In the skewonless **case**, the no-birefringence condition is

$$\chi^{ijkl} = \tfrac{1}{2} (-h)^{1/2}[h^{ik} h^{jl} - h^{il} h^{kj}]\psi + \varphi e^{ijkl}, \qquad (4)$$

(Ni 1983a, 1984). Equation (4) has also been proved more recently without weak-field approximation for **non-**birefringent medium in the skewonless case (Favaro and Bergamin 2011, Lämmerzahl and Hehl 2004). Polarization measurements of electromagnetic waves from pulsars and cosmologically distant astrophysical sources yield stringent constraints agreeing with (4) down to $2\times10^{-32}$ fractionally (for a review, see Ni 2010).

Constraint of the light cone metric $h^{il}$ to matter metric $g^{il}$ up to a scalar factor can be obtained from Hughes-Drever-type experiments and constraint on the dilaton $\psi$ to 1 (constant) can be obtained from Eötvös-type experiments to high precision (Ni 1983a, 1983b, 1984).

We note that in looking for the empirical foundation of EEP above, only the axion field and Type II skewon field are not well constrained. In Section 4, observational constraints on pseudoscalar-photon interaction (axion field) are reviewed. These give an upper limit of a few degrees for the mean value part of the difference of pseudoscalar field at the last scattering surface and at the observation point. In this paper, we look into further constraint/evidence of pseudoscalar-photon interaction in CMB B-mode polarization, especially on the fluctuation/variation part.

The interaction Lagrangian density for the pseudoscalar-photon interaction is

$$L_I^{(EM-Ax)} = -(1/(16\pi)) \varphi\, e^{ijkl} F_{ij} F_{kl} = -(1/(4\pi)) \varphi_{,i}\, e^{ijkl} A_j A_{k,l} \quad (\text{mod div}), \quad (5)$$

where 'mod div' means that the two Lagrangian densities are related by integration by parts in the action integral (Ni 1973, 1974, 1977). If we assume that the $\varphi$-term is local CPT invariant, then $\varphi$ should be a pseudoscalar (function) since $e^{ijkl}$ is a pseudotensor density. Note that sometimes one inserts a constant parameter to this term; here we absorb this parameter into the definition of the pseudoscalar field $\varphi$. The Maxwell equations (Ni 1973, 1977) become

$$F^{ik}{}_{;k} + (-g)^{-1/2} e^{ikml} F_{km} \varphi_{,l} = 0, \quad (6)$$

where the derivation ';' is with respect to the Christoffel connection. The Lorentz force law is the same as in metric theories of gravity or general relativity. Gauge invariance and charge conservation are guaranteed. The modified Maxwell equations (6) are conformal invariant also.

In a local inertial (Lorentz) frame of the $g$-metric, (6) is reduced to

$$F^{ik}{}_{,k} + e^{ikml} F_{km} \varphi_{,l} = 0. \quad (7)$$

Analyzing the wave into Fourier components, imposing the radiation gauge condition, and solving the dispersion eigenvalue problem, we obtain $k = \omega + (n^\mu \varphi_{,\mu} + \varphi_{,0})$ for right circularly polarized wave and $k = \omega - (n^\mu \varphi_{,\mu} + \varphi_{,0})$ for left circularly polarized wave (Ni 1973; see Ni 2010 for a review). Here $n^\mu$ is the unit 3-vector in the propagation direction. The group velocity is independent of polarization:

$$v_g = \partial\omega/\partial k = 1. \quad (8)$$

There is no birefringence. This property is well known (Ni 1973, Ni 1984, Hehl and Obukhov 2003, Itin 2013]. For the right circularly polarized electromagnetic wave, the propagation from a point $P_1$ (4-point) to another point $P_2$ adds a phase of $\alpha = \varphi(P_2) - \varphi(P_1)$ to the wave; for left circularly polarized light, the added phase will be opposite in sign (Ni 1973). Linearly polarized electromagnetic wave is a superposition of circularly polarized waves. Its polarization vector will then rotate by an angle $\alpha$.

When we integrate along light (wave) trajectory in a global situation, the total polarization rotation (relative to no $\varphi$-interaction) is again $\alpha = \Delta\varphi = \varphi(P_2) - \varphi(P_1)$ for $\varphi$ is a scalar field where $\varphi(P_1)$ and $\varphi(P_2)$ are the values of the scalar field at the beginning and end of the wave. When the propagation distance is over a large part of our observed universe, we call this phenomenon cosmic polarization rotation (Ni 2008).

In the CMB polarization observations, the variations and fluctuations due to pseudoscalar-modified propagation can be expressed as $\delta\varphi(P_2) - \delta\varphi(P_1)$, where $\delta\varphi(P_1)$ is the variation/fluctuation at the last scattering surface. $\delta\varphi(P_2)$ at the present observation point (fixed) is zero. Therefore the covariance of fluctuation $<[\delta\varphi(P_2) - \delta\varphi(P_1)]^2>$ is the covariance of $\delta\varphi^2(P_1)$ at the last scattering surface. Since our Universe is isotropic and homogeneous at the last scattering surface to $\sim 10^{-5}$, this covariance is $\sim (\xi \times 10^{-5})^2 \varphi^2(P_1)$ where the parameter $\xi$ depends on various cosmological models (Ni 2008).

In the propagation, E-mode polarization will rotate into B-mode polarization with $\sin^2 2\alpha$ ($\approx 4\alpha^2$ for small $\alpha$) fraction of power. For uniform rotation across the sky, the azimuthal eigenvalue $\ell$ is invariant under polarization rotation and does not change. For small angle,

$$\alpha = \varphi(P_2) - \varphi(P_1) = [\varphi(P_2) - \varphi(P_1)]_{mean} + \delta\varphi(P_1) = <\alpha> + \delta\alpha, \qquad (9)$$

$$\underline{\alpha}^2 \equiv <\alpha^2> = ([\varphi(P_2) - \varphi(P_1)]_{mean})^2 + \delta\varphi^2(P_1) = <\alpha>^2 + \delta\alpha^2, \qquad (10)$$

where $\underline{\alpha} \equiv <\alpha^2>^{1/2}$ is the root mean-square-sum polarization rotation angle, $[\varphi(P_2) - \varphi(P_1)]_{mean} = <\alpha>$ and $\delta\alpha = \delta\varphi(P_1)$.

In translating the power distribution to azimuthal eigenvalue variable $\ell$, we need to insert a factor $\zeta(\ell) \approx \ell$ in front of $\delta_\ell\varphi^2(2)$ to take care of the nonlinear conversion to $\ell$ due to fluctuations. For a uniform rotation with angle $\alpha$ across the sky, the rotation of (original) E-mode power $C_l^{EE}$ into B-mode power $C_l^{BB,obs}$ and $EB$ correlation power are given by (see, e.g., Keating et al. 2013):

$$C_l^{BB,obs} = C_l^{BB} \cos^2(2\alpha) + C_l^{EE} \sin^2(2\alpha), \qquad (11a)$$
$$C_l^{EB,obs} = (C_l^{EE} - C_l^{BB}) \sin(2\alpha) \cos(2\alpha). \qquad (11b)$$

The rotation of (original) B-mode power $C_l^{BB}$ into E-mode power $C_l^{EE,obs}$ and $EB$ correlation power is small and negligible in our analysis since the primordial B-mode is small compared with the E-mode power. For CPR fluctuation in a patch of sky, if we consider only the components deriving from E-mode power $C_l^{EE}$ for $\ell > 20$, the rotated B-mode $\ell$-power spectrum $C_l^{BB,obs}$ and the rotated $EB$ correlation power $C_l^{EB,obs}$ for small CPR angle $\underline{\alpha}$ is accurately given by

$$C_l^{BB,obs} \approx C_l^{EE} \sin^2(2\underline{\alpha}) \approx 4\underline{\alpha}^2 \, C_l^{EE}. \qquad (12a)$$
$$C_l^{EB,obs} \approx C_l^{EE} \sin(2\underline{\alpha}) \cos(2\underline{\alpha}) \approx 2\underline{\alpha} \, C_l^{EE}. \qquad (12b)$$

The accuracy of equation (12a) is shown in Figure 1 which compares the E-mode power spectrum from Figure 10 of Lewis & Challinor (2006) and the CPR B-mode power spectrum from Fig. 5 of Zhao & Li (2014). They are almost identical up to a scale $4\underline{\alpha}^2$ although their input parameters are slightly different.

The present BICEP2 data group about 32 azimuthal eigen-modes into one band with the lowest $\ell$ contribution greater than 20; $\zeta(\ell)$ is virtually equal to one. We will set it to 1 in our analysis. For precise measurement of variations/fluctuations, direct processing of data without first evaluation of the $\ell$ components may be an alternative method.

Some CMB polarization projects apply a uniform angle derotation to the measured Q and U maps, by minimizing the *TB* and *EB* power to compensate for insufficient calibrations of the polarization angle (Keating et al. 2013). This procedure will automatically eliminate the sum of any systematic error in the polarization angle calibration and of any uniform CPR, if it exists. If calibration errors of the polarization angle were small compared to uniform CPR, in principle this minimization procedure could provide an estimate of the uniform CPR angle <α>. However, since the systematic errors in the polarization angle are of the same order of current upper limits to uniform CPR, in fact this procedure is equivalent to assume no uniform CPR and will preclude any information on it.

SPTpol first estimates the lensing potential from a Herschel-SPERE map of the cosmic infrared background and constructs a template for the lensing B-mode signal by combining SPTpol measured E-mode polarization with estimated lensing potential. SPTpol then compares this constructed template to its directly measured B-modes to determine the lensing B-mode by correlation method. In this way the uniform (mean) CPR would not be included. The CPR fluctuations incurred at the lensing site would be included and correlated with the pseudoscalar fluctuations at the lensing site (z ~ 2-5). Assuming that the linear perturbation scheme works, the uniform (mean) part of the pseudoscalar field at lensing is developed from the uniform (mean) part of the pseudoscalar field at the last scattering surface and the pseudoscalar fluctuations at lensing are developed also from the fluctuations at the last scattering surface (z ~

1100). In most theoretical models, the pseudoscalar fluctuations are correlated to the density fluctuations at the last scattering surface. Therefore the pseudoscalar field fluctuations at lensing would also be correlated with density fluctuations at lensing. The strength of this correlation depends on cosmological models with pseudoscalar field. Effects from re-ionization are minor but could also be included. Therefore in the fits with SPTpol data, we include a percentage correlation parameter κ (this parameter could also be greater than one in some models).

## 3. MODELING THE DATA

In this section, we model the available data for the BB power spectrum with the three effects mentioned in the Introduction. The theoretical spectrum of the inflationary gravitational waves and the lensing contribution to B-mode are extracted from the BICEP2 paper (Ade et al. 2014b). The power spectrum $C_l^{BB,\text{obs}}$, induced by any existing CPR angle (eq. 12a), is obtained from the theoretical E-mode power spectrum $C_l^{EE}$ of Lewis & Challinor (2006) and is shown in Figure 1.

The data on the BB power spectrum are the 9 points of BICEP2 in Fig. 14 of Ade et al. (2014b) for the low multipole part ($21 \leq \ell \leq 335$), the 4 points of SPTpol in Fig. 2 of Hanson et al. (2013) for $200 \leq \ell \leq 3000$, and the 4 points of POLARBEAR (Ade et al. 2014c) for $500 \leq \ell \leq 2100$, based on observations at 150 GHz on three regions of the sky for a total of 30 square deg. Although the S/N ratio of the recently published results of POLARBEAR for the CMB B-mode polarization is considerably lower than that of the one obtained by the SPTpol collaboration in the same range of $\ell$, the SPTpol data, depending on theoretical models, only include CPR partially in their lensing correlation measurement. Therefore we have considered both data sets in our fits. As it turned out, the three experiments give similar constraints on the relevant CPR parameters.

Since a uniform derotation is implemented by both the BICEP2 and POLARBEAR experiments, by minimizing EB and TB power, in order to compensate for the relatively large errors in the calibration of the polarization angle, their data can only give constraints on the CPR fluctuations (variance), but not on the CPR mean angle. SPTpol have not applied such derotation, although their systematic uncertainty on the polarization angle is about 1° at 150 GHz, but they give the cross-correlation of their measured B-modes with the lensing B-modes inferred from CIB fluctuations as measured by Herschel, rather than the BB autocorrelation. Therefore also from the SPTpol data it is possible to derive a constraint only on the CPR fluctuations, not on the mean angle.

Table 1 and Figures 2-7 show the results of our fits for various combinations of the available data and for different values of the correlation percentage κ of the pseudoscalar field fluctuations with the density fluctuations at lensing in the SPTpol case. Figure 2 shows the results of fitting of the CPR fluctuation and scalar-to-tensor ratio for BICEP2. Figure 3 shows the results of fitting of the CPR fluctuation to the POLARBEAR experiment. Figure 4 shows the results of joint fitting to BICEP2 and POLARBEAR. Figure 5 for the results of fitting to SPTpol experiment. Figure 6 for joint BICEP2-SPTpol fitting and Figure 7 for joint BICEP2-POLARBEAR-SPTpol fitting. Figure 6(b) shows how the $\chi^2$ and the CPR angle fluctuations vary with the CPR-SPTpol correlation percentage κ, considering BICEP2 and SPTpol data: the minimum $\chi^2$ is obtained at 70% (but it is a shallow minimum) and the largest CPR fluctuations are obtained at 23%. Figure 7(b) show the same for all three experiments: in this case the minimum $\chi^2$ is obtained at 86% and the largest CPR variance is at 29%.

Table 1. Results of fitting the CPR fluctuation $\delta\alpha^2$ and/or the scalar-to-tensor ration r to BICEP2 (Ade et al. 2014b), POLARBEAR (Ade et al. 2014c), SPTpol (Hanson et al. 2013) and their joint combinations.

| Experiment | Fitting Parameter | | $\chi_{min}^2$ (# of data points − # of fitting parameters) | 1 σ upper limit on CPR fluctuation amplitude $<\delta\alpha^2>^{1/2}$ [mrad] |
|---|---|---|---|---|
| | $<\delta\alpha^2>$ [mrad$^2$] | r | | |
| BICEP2 | 339±455 | 0.196±0.033 | 10.424 (9 – 2) | 28.2 (1.61°) |
| POLARBEAR | 89±535 | --[a] | 3.73 (4 – 1) | 25.0 (1.43°) |
| POLARBEAR + BICEP2 | 265±397 | 0.198±0.033 | 14.31 (13 – 2) | 25.7 (1.47°) |
| SPTpol | $\kappa^{-1}$(233±148) | --[a] | 2.61 (4 – 1) | $\kappa^{-1}$19.5 ($\kappa^{-1}$1.12°) |
| SPTpol (23%) + BICEP2 | 486±411 | 0.190±0.033 | 13.81 (13 – 2) | 29.9 (1.72°) |
| SPTpol (70%) + BICEP2 | 340±270 | 0.196±0.033 | 13.05 (13 – 2) | 24.7 (1.42°) |
| SPTpol (100%) + BICEP2 | 244±202 | 0.199±0.033 | 13.13 (13 – 2) | 21.1 (1.21°) |
| POLARBEAR + BICEP2 + SPTpol (29%) | 392±353 | 0.194±0.033 | 19.27 (17 – 2) | 27.3 (1.56°) |
| POLARBEAR + BICEP2 + SPTpol (86%) | 264±218 | 0.198±0.033 | 18.34 (17 – 2) | 22.0 (1.26°) |

[a] r is set to 0.2 to conform to BICEP2 data; the effect of setting r to 0.2 or 0 to the CPR fluctuation fitting is small since the power of non-vanishing r contribution to the total power is small for the multipoles measured in the POLARBEAR and SPTpol experiments.

## 4. CONSTRAINTS ON COSMIC POLARIZATION ROTATION

The CPR has not yet been detected. Upper limits have been obtained from radio galaxies polarization, both in the radio and in the optical/UV (di Serego Alighieri et al. 2010), and from CMB polarization anisotropies. These limits have been reviewed by

di Serego Alighieri (2011): all methods have reached an accuracy of about 1° and 3σ upper limits to any rotation of a few degrees. Since this review only minor revisions of the limits from the CMB have appeared (see Table 2). Gluscevic et al. (2012) have searched for direction-dependent CPR with WMAP 7-year data and obtain an upper limit on the rms rotation angle $<\alpha^2>^{1/2}<9.5°$ for $0<\ell<512$ and an upper limit at 68% CL of about 1° on the quadrupole of a scale-independent rotation-angle power spectrum.

Table 2: Recent constraints on uniform CPR angle from CMB E-mode polarization

| Experiment | Frequencies | $\ell$-range | CPR angle | Reference |
| --- | --- | --- | --- | --- |
| WMAP9 | 41,61,94 GHz | 2-800 | -0.36°±1.24°±1.5° | Hinshaw et al. 2012 |
| BICEP1 | 100,150 GHz | 20-335 | -2.77°±0.86°±1.3° | Kaufman et al. 2014 |

The rotation angle given by the revision of BICEP1 data by Kaufman et al. (2014) formally corresponds to a 1.78 sigma detection of CPR. However, taking into account the uncertainties involved in the calibration of the BICEP1 polarization angle, the problems in correcting for galactic Faraday rotation, (Section VI.B of Kaufman et al. 2014), and the inconsistency of this result with the one of QUaD, for which Brown et al. (2009) give a CPR angle of 0.64°±0.5°(stat.)±0.5°(syst.), we do not consider this CPR "detection" as final. Recently Gubitosi and Paci (2013) have reviewed the constraints on CPR, but they have considered only those deriving from CMB polarization data.

By modeling the SPTpol, POLARBEAR and BICEP2 data on the B-mode polarization power spectrum, taking into account the inflationary gravitational waves, the gravitational lensing, and the component induced by CPR, as explained in the previous sections, we can derive a new constraint on CPR fluctuations, since the present data limit $<\delta\alpha^2>$. In fact in our case, since we cannot predict which should be the correlation percentage between lensing E-mode and B-mode CPR effects, the most conservative CPR constraint is set at the percentage which gives the maximum $<\delta\alpha^2>$, i.e. 29% (Fig. 7b and upper part of Fig. 7a). In fact, we could not consider instead the correlation percentage giving the lowest $\chi^2$, i.e. 86% (Fig. 7b and lower part of Fig. 7a), also because all correlation percentages between 30% and 100% give $\chi^2$ values within 10% of the best one, hence they all are statistically acceptable. Therefore, using the data from the three available experiments, we can set a limit: $<\delta\alpha^2> \leq 392+353=745$ mrad$^2$, i.e. $<\delta\alpha^2>^{1/2} \leq 27.3$ mrad or 1.56° at 68% C.L.

Our models give a value for r which is slightly lower than the value given by the BICEP2 collaboration (Ade et al. 2014b), but still in disagreement with the Planck 6-parameter-fit limit (Ade et al. 2014a). However the Planck limit is relaxed to r<0.26 when running is allowed with $dn_s/d\ln k = -0.022\pm0.010$ (Ade et al. 2014b). Our fitted r-value gives a 6σ detection, which agrees with the BICEP2 results and shows robustness to adding CPR effects.

Lee et al. (2014) have recently modeled the power spectrum of the B-mode CMB polarization, including a component deriving from CPR (they call it "cosmological birefringence"), therefore decreasing the "primordial" component and bring it within the limit r<0.11 set by Ade et al. (2014a). However they have considered only the BICEP2 data (Ade et al. 2014b) at low $\ell$, not the SPTpol and POLARBEAR data (Hanson et al. 2014 and Ade et al. 2014c) at high $\ell$, which together better constrain the CPR. Therefore they allow for a CPR component much larger than we do. We believe that our approach is a considerable improvement.

## 5. DISCUSSION AND OUTLOOK

In this paper we have investigated, both theoretically and experimentally, the possibility to detect CPR, or set new constraints to it, using its coupling with the B-mode power spectra of the CMB. Three experiments have detected B-mode polarization in the CMB: SPTpol (Hanson et al. 2013) for $200 \le \ell \le 3000$, BICEP2 (Ade et al. 2014b) for $21 \le \ell \le 335$, and POLARBEAR (Ade et al. 2014c) for $500 \le \ell \le 2100$.

The practical realization of our suggestion has a problem. As explained in Section 3, currently available data on B-mode CMB polarization only allow to constrain the fluctuations of the CPR angle, not its mean value. Using all the available data we set a constraint to the fluctuations of the CPR: $<\delta\alpha^2>^{1/2} \le 27.3$ mrad (1.56°) at 68% C.L.

However we believe that the method of investigating the CPR from data on the B-mode CMB polarization will become useful with future experiments which will solve the problems of the calibration of the polarization angle. For example Naess et al. (2014) have very recently reported about a new measurement of the polarization of the Crab Nebula, an important calibration source, at 146 GHz with the Atacama Cosmology Telescope, giving an accuracy of 0.5° on the angle, an improvement over previous measurements. Concerning the prospects for improvements in the detection/constraints on CPR in the near future, the Planck satellite (http://www.rssd.esa.int/index.php?project=Planck) is expected to allow for a considerable step forward, since it will have a sensitivity to polarization rotation of the order of a tenth of a degree and systematic errors of the same order (Gruppuso et al. 2012), provided that calibration procedures of sufficient accuracy for the polarization orientation are implemented. In any case the Planck polarization results are foreseen only towards the end of this year. Also the Keck Array, which is a set of 5 telescopes very similar to the BICEP2 one, is expected to bring considerable improvements on current CMB polarization measurements. Three of these telescopes are already operating and results are expected soon, particularly for B-mode polarization due to inflationary gravitational waves (Pryke et al. 2013).

As far as we are aware, there is no current plan for extending the optical/UV polarization measurement of distant radio galaxies to a level which would bring considerable improvements on the CPR constraints obtained so far from these objects. However it would be desirable to perform detailed spatially-resolved observations of the optical/UV polarization of a few radio galaxies, of the kind performed on 3C 265 (Wardle, Perley & Cohen 1997) and selected to be in the sky areas also observed by CMB polarization experiments on the ground. In fact these observations are likely to constrain the CPR angle with an accuracy of better than 1°, thereby providing a better calibration for the absolute polarization angle than available so far to these experiments.

If pseudoscalar-photon interactions exist, a natural cosmic variation of the pseudoscalar field at the decoupling era is $10^{-5}$ fractionally. The CPR fluctuation is then of the order of $10^{-5}\varphi_{decoupling\text{-}era}$. We will look for its possibility of detection or more constraints in future experiments.

Acknowledgements

We thank Brian Keating and the referee for useful comments, and Matteo Galaverni for useful discussion. One of us (WTN) would also like to thank the National Science Council (Grants No. NSC101-2112-M-007-007 and No. NSC102-2112-M-007-019) and the National Center for Theoretical Sciences (NCTS) for supporting this work in part.

REFERENCES

Ade, P.A.R. et al. (Planck Collaboration) 2014a, A&A, in press, also arXiv:1303.5076
Ade, P.A.R. et al. (BICEP2 Collaboration) 2014b, Phys. Rev. Lett., 112，241101
Ade, P.A.R. et al. (POLARBEAR Collaboration) 2014c, arXiv:1403.2369
Brown, M.L. et al. (QUaD Collaboration) 2009, ApJ 705, 978
di Serego Alighieri, S., Finelli, F. & Galaverni, M. 2010, ApJ 715, 33

di Serego Alighieri, S. 2011, in *From Varying Couplings to Fundamental Physics,* C. Martins and P. Molaro (eds.), ASSP, Springer, p. 139, also arXiv:1011.4865

Favaro, A. & Bergamin, L. 2011, Annalen der Physik 523, 383

Fixsen, D. J. 2009, Ap. J. 707, 916

Gluscevic, V., Hanson, D., Kamionkowski, M. & Hirata, C.M. 2012, Phys. Rev. D 86, 103529

Gruppuso, A., Natoli, P., Mandolesi, N., De Rosa, A., Finelli, F. & Paci, F. 2012, JCAP 2, 23

Gubitosi, G. & Paci, F. 2013, JCAP 2, 20

Hanson, D. et al. (SPTpole Collaboration) 2013, Phys. Rev. Lett. 111, 141301

Hehl, F.W., Yu, N., and Obukhov, G.F. 2003, *Foundations of Classical Electrodynamics: Charge, Flux, and Metric* (Birkhäuser: Boston, MA)

Hinshaw, G. et al. (WMAP Collaboration) 2013, ApJS 208, 19

Holder, G.P. et al. (SPT Collaboration) 2013, ApJL, 771, L16

Itin, Y. 2013, Phys. Rev. D, 88, 107502

Komatsu, E. et al. (WMAP Collaboration) 2011, ApJS 192, 18

Kaufman, J.P. et al. (BICEP1 Collaboration) 2014, Phys. Rev. D 89, 062006

Keating, B.G., Shimon, M. & Yadav, A.P.S. 2013, ApJL, 762, L23

Lämmerzahl, C. & Hehl, F.W. 2004, Phys. Rev. D, 70, 105022

Lee, S., Liu, G.-C. & Ng, K.-W. 2014, arXiv:1403.5585

Lewis, A. & Challinor, A. 2006, Phys. Rept., 429, 1

Naess, S. et al. 2014, arXiv:1405.5524

Ni, W.-T. 1973, A Nonmetric Theory of Gravity, Montana State University, http://astrod.wikispaces.com/

Ni, W.-T. 1974, Bull. Am. Phys. Soc., 19, 655

Ni, W.-T. 1977, Phys. Rev. Lett., 38, 301

Ni, W.-T. 1983a, in *Proceedings of the 1983 International School and Symposium on Precision Measurement and Gravity Experiment*, ed. W.-T. Ni, (National Tsing Hua University, Hsinchu, Taiwan, Republic of China), 491, http://astrod.wikispaces.com/

Ni, W.-T. 1983b, in *Proceedings of the 1983 International School and Symposium on Precision Measurement and Gravity Experiment*, ed. W.-T. Ni (National Tsing Hua University, Hsinchu, Taiwan, Republic of China), 519, http://astrod.wikispaces.com/

Ni, W.-T. 1984, in *Precision Measurement and Fundamental Constants II*, ed. B. N. Taylor & W. D. Phillips, Natl. Bur. Stand. Spec. Publ. 617, 647

Ni, W.-T. 2008, *Prog. Theor. Phys. Suppl.,* 172, 49

Ni, W.-T. 2010, *Reports on Progress in Physics,* 73, 056901

Ni, W.-T. 2014, Phys. Lett. A 378, 1217

Polnarev, A. G. 1985, SvA 29, 607

Pryke, C. et al. (BICEP2 and Keck-Array Collaborations) 2013, in *Astrophysics from Antartica*, IAU Symp. S288, Proc. of the IAU, Vol. 8, p. 68

Wardle, J.F.C., Perley, R.A. & Cohen, M.H. 1997, Phys. Rev. Lett. 79, 1801

Zaldarriaga, M. & Seljak, U. 1997, Phys. Rev. D, 55, 18309

Zhao, W., & Li, M. 2014, Phys. Rev. D, 89, 103518

Figures

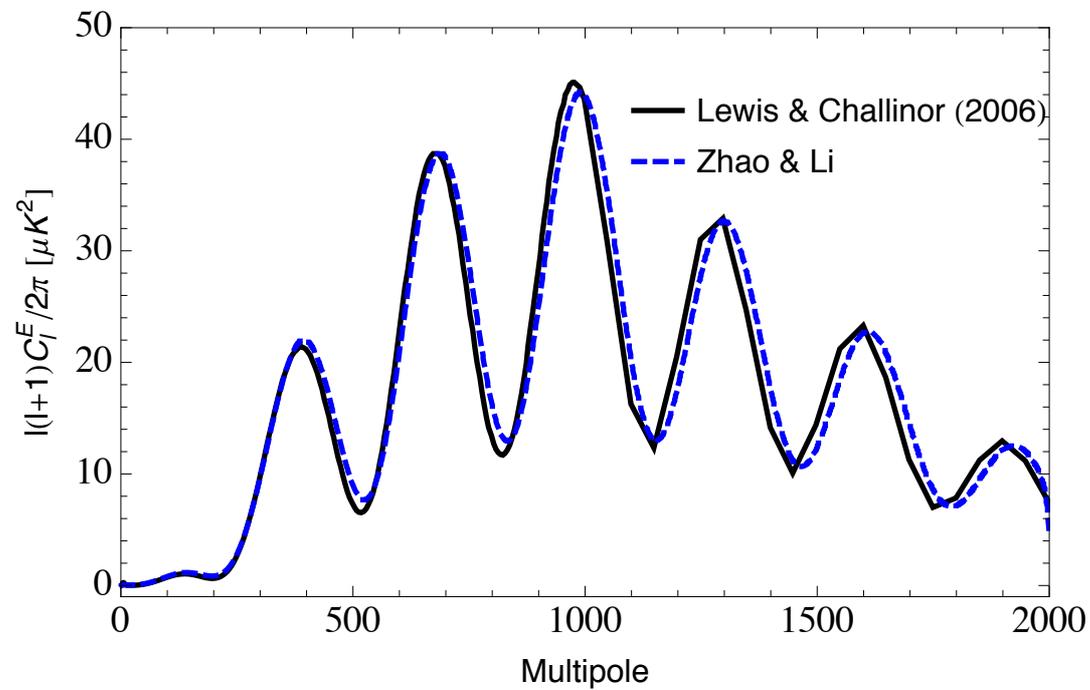

Figure 1: Comparison of the E-mode power spectrum from Figure 10 of Lewis & Challinor (2006) and the CPR B-mode power spectrum from Fig. 5 of Zhao & Li (2014) up to a scale $4\underline{\alpha}^2$.

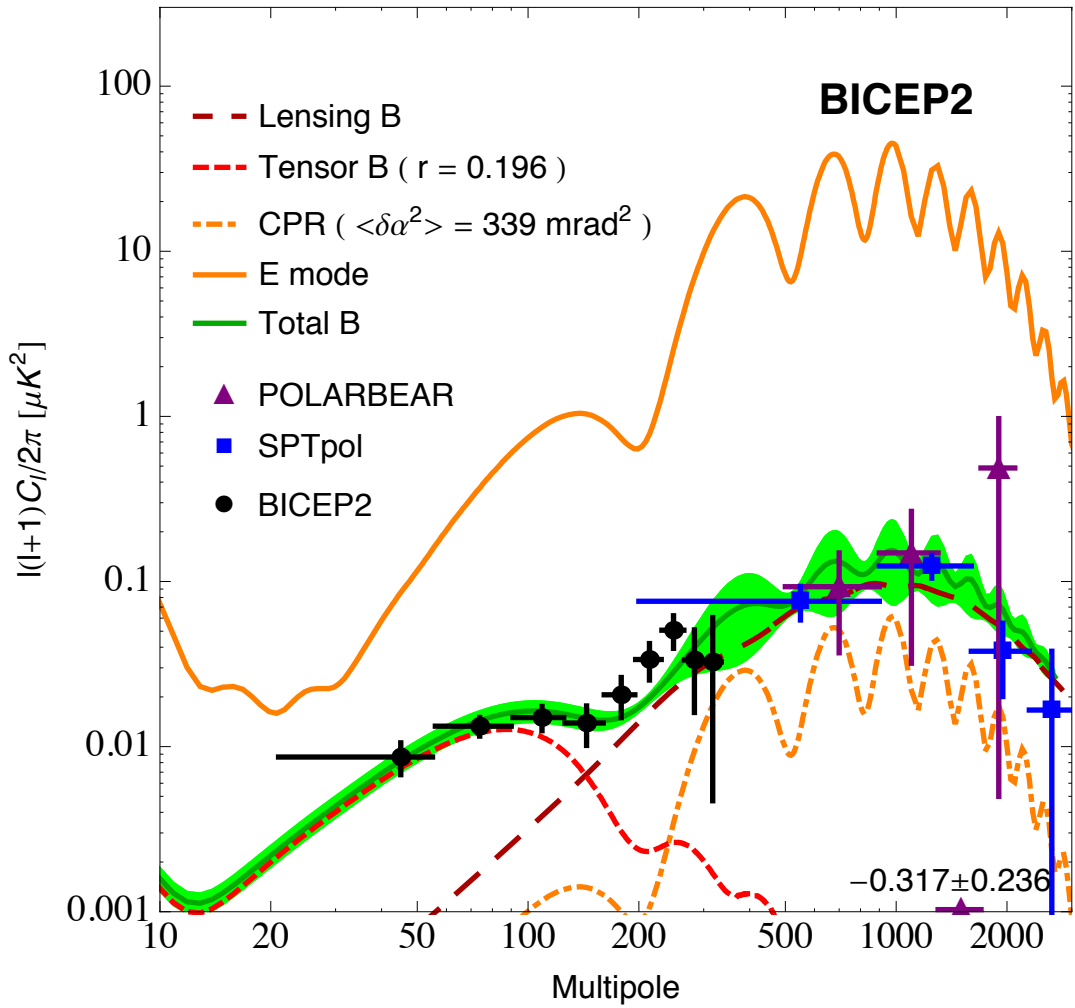

Figure 2. The B-mode spectrum showing the best fit (dark green line) and the one σ region (green band) to the 9 BICEP2 data points (black filled circles) with the vertical bar showing standard deviation of each data point and the horizontal bar showing the binning interval. The E-mode is plotted for reference. The power of the second highest multipole band of POLARBEAR (l from 1300 to 1700) is negative, i.e. $-0.317\pm0.236$ μK$^2$; we show the binning interval on the horizontal axis with the data value in Arabic numerals above the binning interval.

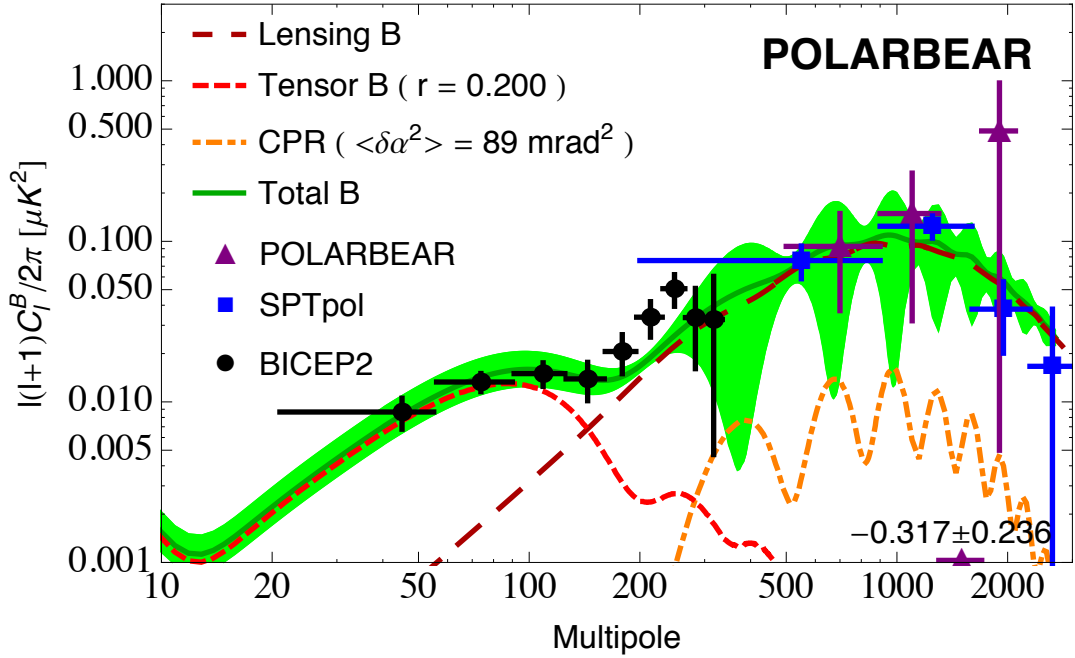

Fig.3. Same as Fig. 2, but for the POLARBEAR data points (purple filled triangle). r is set to 0.2 to conform to BICEP2 data; the effect of setting r to 0.2 or to 0 for the fitting of CPR fluctuation is small since the contributed power of non-vanishing r to the total power is small for the multipoles measured in the POLARBEAR experiment.

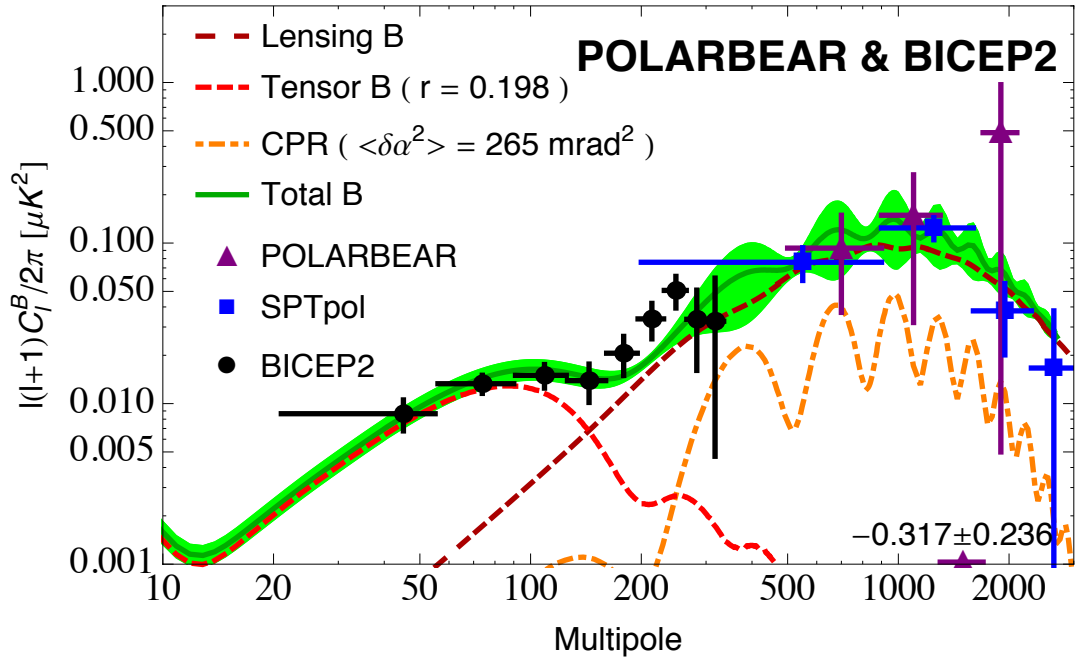

Fig. 4. Same as Fig. 2, but for the BICEP2 (black filled circles) and POLARBEAR data points (purple filled triangle).

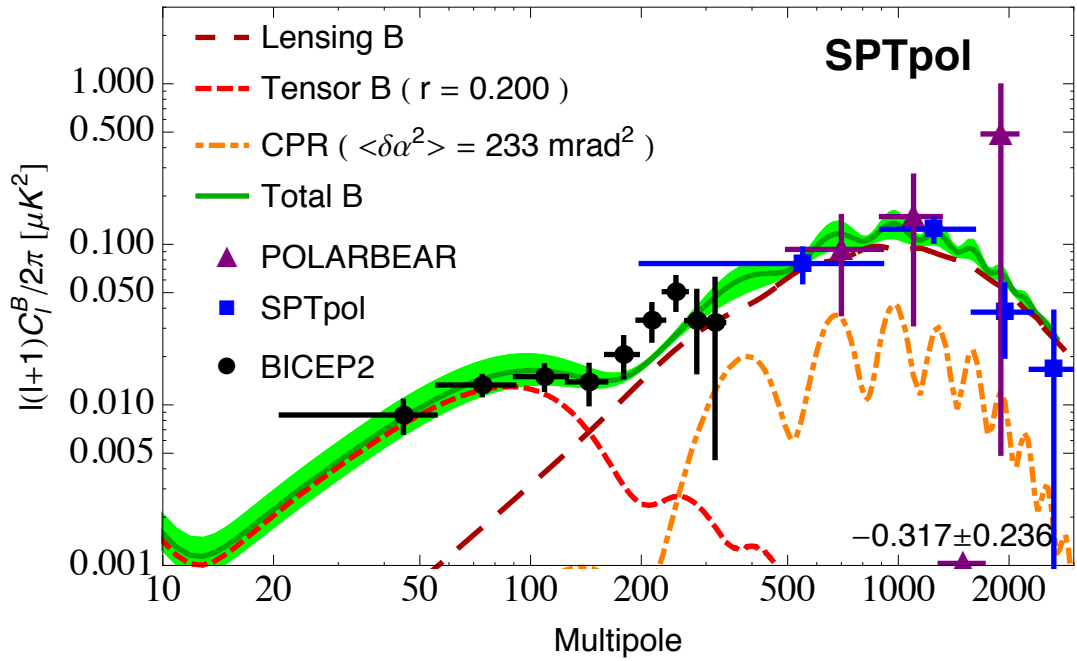

Fig. 5 Same as Fig. 2, but for the SPTpol data points (blue filled square) with 100 % CPR-SPTpol correlation. This is a one-parameter fit to the CPR effect. For theoretical models with CPR-SPTpol correlation $\kappa$, the fitted value of $<\delta\alpha^2>$ is $\kappa^{-1}(233\pm148)$ mrad$^2$ instead. r is set to 0.2 to conform to BICEP2 data; the effect of setting r to 0.2 or to 0 for the fitting of CPR fluctuation is small since the power contribution of non-vanishing r to the total power is small for the multipoles measured in the SPTpol experiment.

(a)

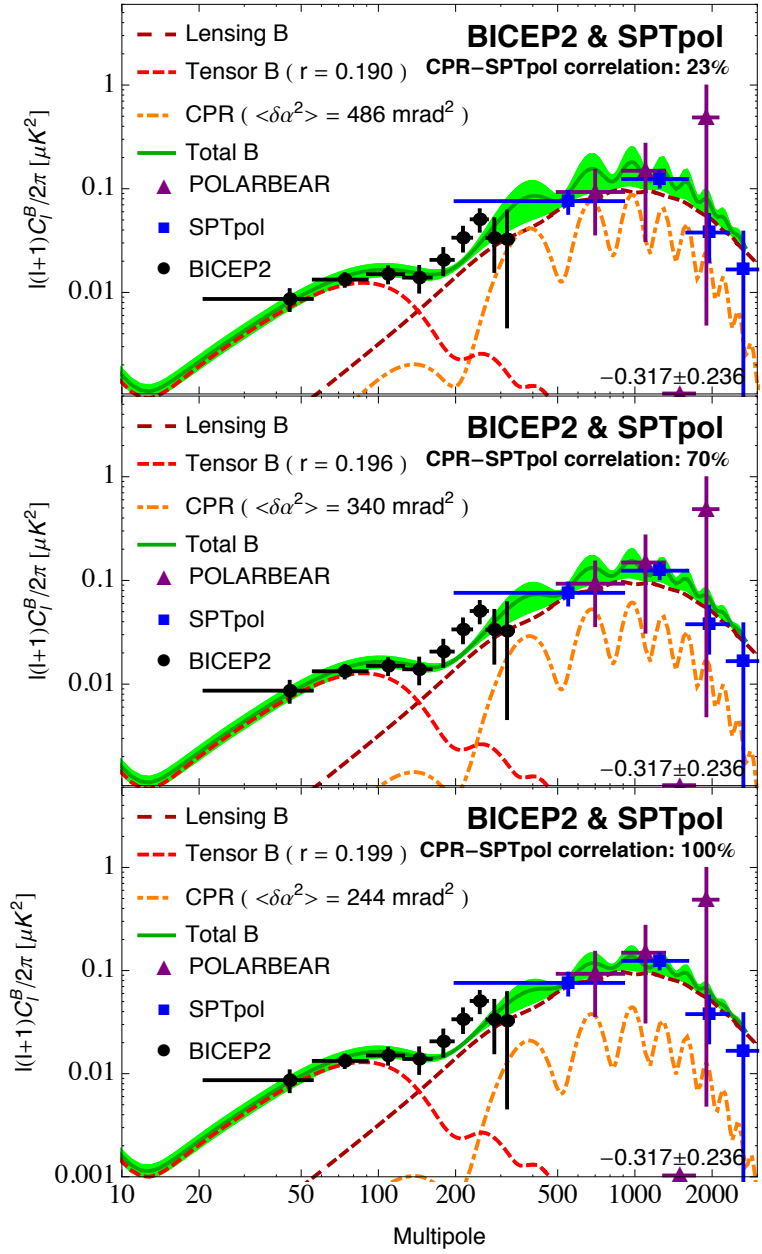

(b)

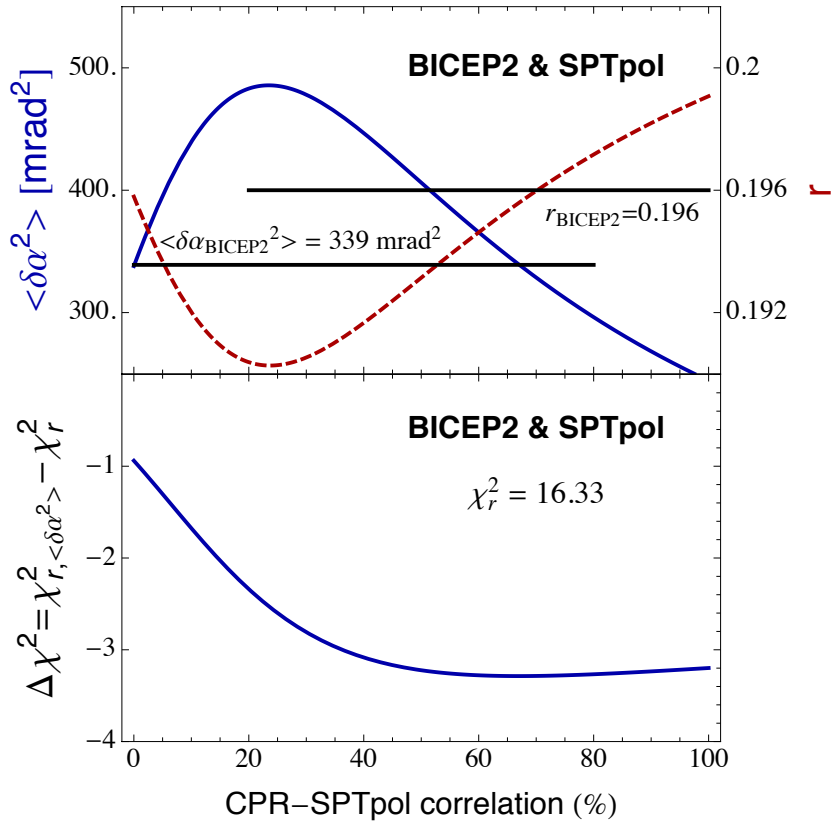

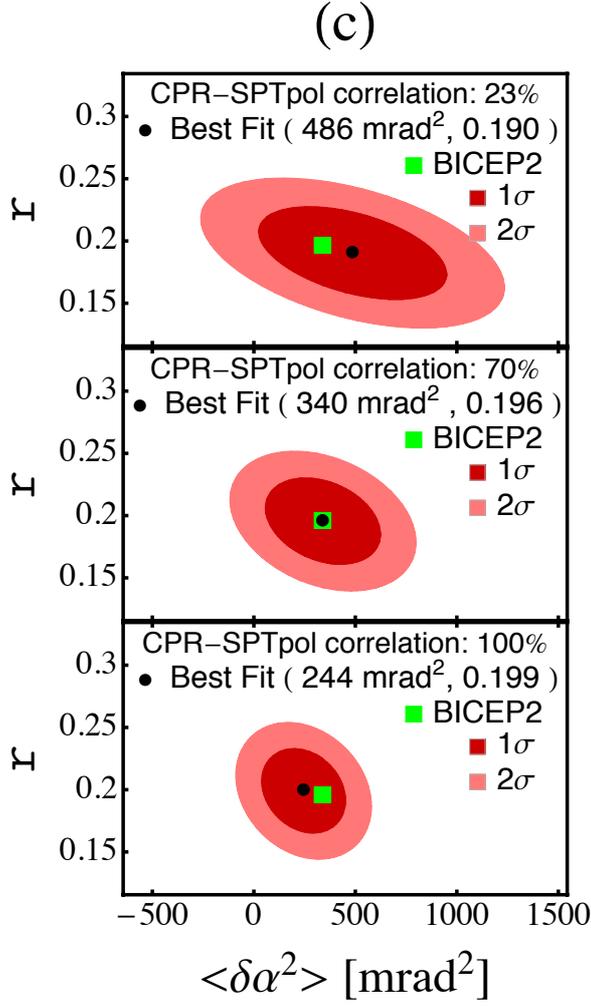

Figure 6. (a) is the same as Fig. 5, but for the BICEP2 data points (black filled circles) and SPTpol data points (blue filled square) with 23 %, 70% and 100 % CPR-SPTpol correlations for the upper part, the middle part and the lower part of figure respectively. (b) shows the dependence of $<\alpha>^2$, r and $\chi^2$, on the CPR-SPTpol correlation κ; (c) shows the 1σ and 2σ contours of the joint constraint on the tensor-to-scalar ratio r and the root-mean-square-sum of CPR angle due to pseudoscalar-photon interaction for three cases in (a).

(a)

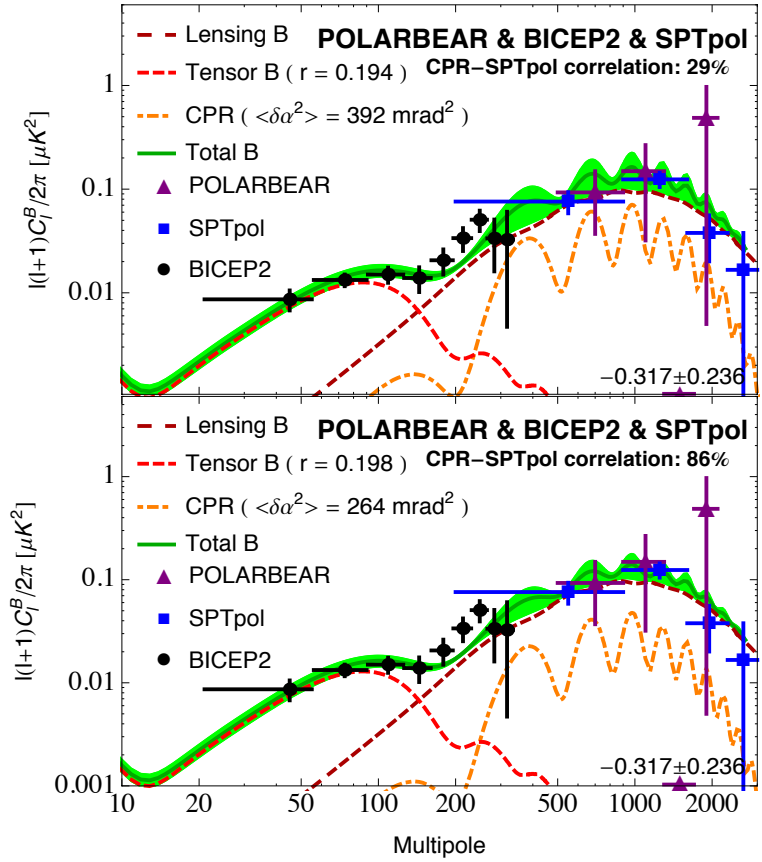

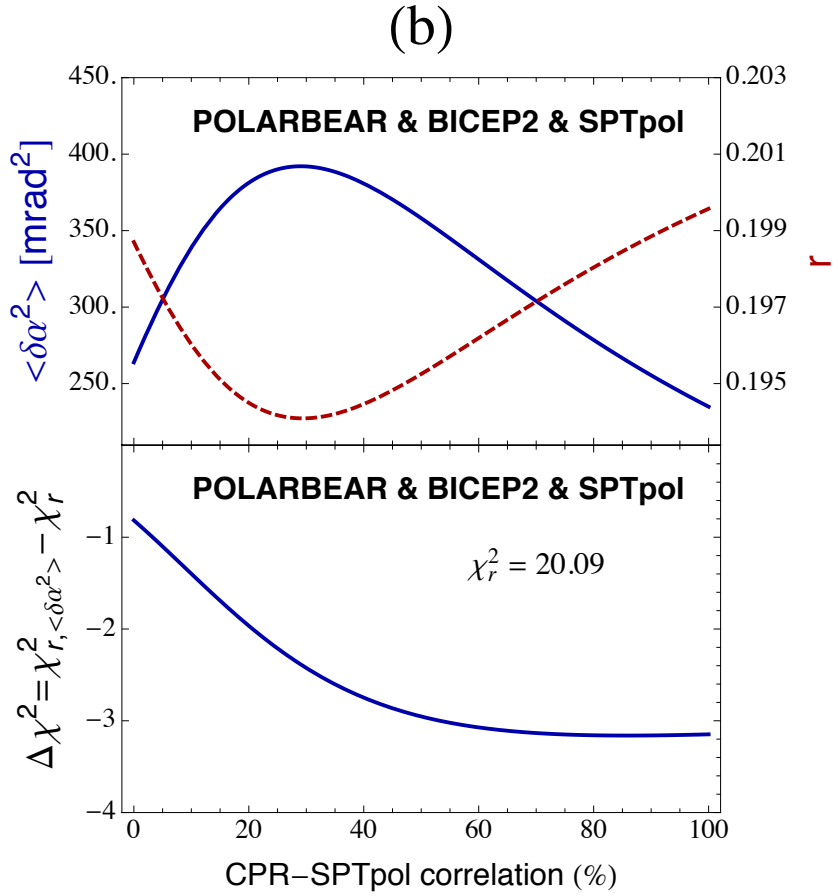

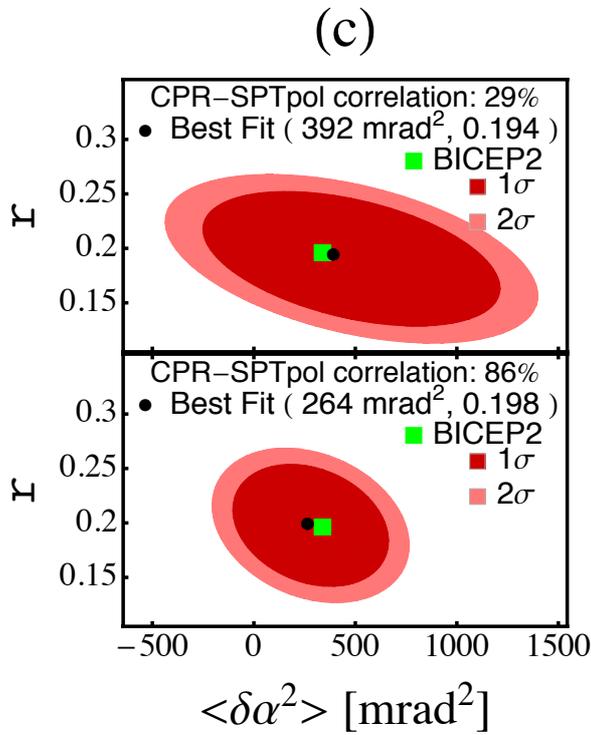

Figure 7. (a) is the same as Fig. 5, but for the BICEP2 data points (black filled circles), POLARBEAR data points (purple filled triangle), and SPTpol data points (purple

filled triangle) with 29 % and 86% CPR-SPTpol correlations for the upper part and the lower part of figure respectively. (b) shows the dependence of $<\alpha>^2$, r and $\chi^2$, on the CPR-SPTpol correlation κ; (c) shows the 1σ and 2σ contours of the joint constraint on the tensor-to-scalar ratio r and the root-mean-square-sum of CPR angle due to pseudoscalar-photon interaction for the two cases (a).